\PassOptionsToPackage{unicode}{hyperref}
\PassOptionsToPackage{hyphens}{url}
\PassOptionsToPackage{dvipsnames,svgnames,x11names}{xcolor}
\documentclass[
  12pt]{article}

  \usepackage{amsthm}
\usepackage{graphicx,psfrag,epsf}
\usepackage{enumerate}
\usepackage{booktabs}
\usepackage{rotating}

\usepackage{amsmath,amssymb}
\usepackage{iftex}
\ifPDFTeX
  \usepackage[T1]{fontenc}
  \usepackage[utf8]{inputenc}
  \usepackage{textcomp} 
\else 
  \usepackage{unicode-math}
  \defaultfontfeatures{Scale=MatchLowercase}
  \defaultfontfeatures[\rmfamily]{Ligatures=TeX,Scale=1}
\fi
\usepackage{lmodern}
\ifPDFTeX\else  
\fi
\IfFileExists{upquote.sty}{\usepackage{upquote}}{}
\IfFileExists{microtype.sty}{
  \usepackage[]{microtype}
  \UseMicrotypeSet[protrusion]{basicmath} 
}{}
\makeatletter
\@ifundefined{KOMAClassName}{
  \IfFileExists{parskip.sty}{%
    \usepackage{parskip}
  }{
    \setlength{\parindent}{0pt}
    \setlength{\parskip}{6pt plus 2pt minus 1pt}}
}{
  \KOMAoptions{parskip=half}}
\makeatother
\usepackage{xcolor}
\makeatletter
\ifx\paragraph\undefined\else
  \let\oldparagraph\paragraph
  \renewcommand{\paragraph}{
    \@ifstar
      \xxxParagraphStar
      \xxxParagraphNoStar
  }
  \newcommand{\xxxParagraphStar}[1]{\oldparagraph*{#1}\mbox{}}
  \newcommand{\xxxParagraphNoStar}[1]{\oldparagraph{#1}\mbox{}}
\fi
\ifx\subparagraph\undefined\else
  \let\oldsubparagraph\subparagraph
  \renewcommand{\subparagraph}{
    \@ifstar
      \xxxSubParagraphStar
      \xxxSubParagraphNoStar
  }
  \newcommand{\xxxSubParagraphStar}[1]{\oldsubparagraph*{#1}\mbox{}}
  \newcommand{\xxxSubParagraphNoStar}[1]{\oldsubparagraph{#1}\mbox{}}
\fi
\makeatother

\usepackage{longtable,booktabs,array}
\usepackage{calc} 
\usepackage{etoolbox}
\makeatletter
\patchcmd\longtable{\par}{\if@noskipsec\mbox{}\fi\par}{}{}
\makeatother
\IfFileExists{footnotehyper.sty}{\usepackage{footnotehyper}}{\usepackage{footnote}}
\makesavenoteenv{longtable}
\usepackage{graphicx}
\makeatletter
\def\maxwidth{\ifdim\Gin@nat@width>\linewidth\linewidth\else\Gin@nat@width\fi}
\def\maxheight{\ifdim\Gin@nat@height>\textheight\textheight\else\Gin@nat@height\fi}
\makeatother
\setkeys{Gin}{width=\maxwidth,height=\maxheight,keepaspectratio}
\makeatletter
\def\fps@figure{htbp}
\makeatother

\makeatletter
\@ifpackageloaded{caption}{}{\usepackage{caption}}
\AtBeginDocument{%
\ifdefined\contentsname
  \renewcommand*\contentsname{Table of contents}
\else
  \newcommand\contentsname{Table of contents}
\fi
\ifdefined\listfigurename
  \renewcommand*\listfigurename{List of Figures}
\else
  \newcommand\listfigurename{List of Figures}
\fi
\ifdefined\listtablename
  \renewcommand*\listtablename{List of Tables}
\else
  \newcommand\listtablename{List of Tables}
\fi
\ifdefined\figurename
  \renewcommand*\figurename{Figure}
\else
  \newcommand\figurename{Figure}
\fi
\ifdefined\tablename
  \renewcommand*\tablename{Table}
\else
  \newcommand\tablename{Table}
\fi
}
\@ifpackageloaded{float}{}{\usepackage{float}}
\floatstyle{ruled}
\@ifundefined{c@chapter}{\newfloat{codelisting}{h}{lop}}{\newfloat{codelisting}{h}{lop}[chapter]}
\floatname{codelisting}{Listing}

\makeatother
\makeatletter
\@ifpackageloaded{caption}{}{\usepackage{caption}}
\@ifpackageloaded{subcaption}{}{\usepackage{subcaption}}
\makeatother

\ifLuaTeX
  \usepackage{selnolig}  
\fi
\usepackage[]{natbib}
\usepackage{bookmark}

\IfFileExists{xurl.sty}{\usepackage{xurl}}{} 
\hypersetup{
  pdftitle={Title},
  pdfauthor={Author 1; Author 2},
  pdfkeywords={3 to 6 keywords, that do not appear in the title},
  colorlinks=true,
  linkcolor={blue},
  filecolor={Maroon},
  citecolor={Blue},
  urlcolor={Blue},
  pdfcreator={LaTeX via pandoc}}

\newcommand{\anon}{1}

\begin{document}

\def\spacingset#1{\renewcommand{\baselinestretch}%
{#1}\small\normalsize} \spacingset{1}


\if1\anon
{
  \title{\bf Two-stage Adaptive Design for Cluster Randomised Trials}
  \author{Samuel I. Watson\thanks{
    The authors gratefully acknowledge funding from the National Institute of Health and Care Research (NIHR; NIHR172073)} and James Martin\hspace{.2cm}\\
    Department of Applied Health Sciences, University of Birmingham}
  \maketitle
} \fi

\if0\anon
{
  \bigskip
  \bigskip
  \bigskip
  \begin{center}
    {\LARGE\bf Two-stage Adaptive Design for Cluster Randomised Trials}
\end{center}
  \medskip
} \fi

\bigskip
\begin{abstract}
Adaptive sample size re-estimation, early stopping, and trial re-design at interim analyses can reduce expected sample sizes in randomised trials. Cluster randomised trials, in which groups of participants are randomly allocated to treatment status, may particularly benefit as they can be costly and their required sample sizes depend on one or more auxiliary parameters governing correlations within and between clusters, which are often estimated with high uncertainty. We adapt a combination test approach to the cluster trial setting allowing for early stopping for futility or efficacy and accounting for correlations between trial stages and other nuisance parameters. We consider design decisions for multi-dimensional sample sizes involving clusters, participants, and time and allowing for modifications to intervention roll-out patterns. We use a Pareto optimality approach to balance objectives relating to different components of the sample size and costs. We also examine the interim estimation of auxiliary parameters and trial re-design for efficiency. We illustrate the methods including an example of a parallel cluster trial re-design and a re-analysis of the large cluster randomised trial E-MOTIVE.  
\end{abstract}

\noindent%
\vfill

\newpage
\spacingset{1.45} 
\section{Introduction}
\label{sec:intro}

Many experiments are designed before any outcomes are observed. The sample size, allocation of experimental units, duration of follow-up, and analysis are fixed in advance. This approach is simple and transparent, but it requires investigators to make design decisions using uncertain information about treatment effects, outcome variability, and other parameters that determine statistical efficiency. Adaptive designs instead allow pre-specified features of an experiment to be modified in response to accumulating data. Common adaptations include stopping early for efficacy or futility, increasing or decreasing the remaining sample size, and modifying future treatment allocation. When appropriately designed, these changes can improve efficiency while retaining control of operating characteristics such as the type I error rate. Adaptive methods are now well established for individually randomised clinical trials, with an extensive methodological literature \citep{Jennison2000,Bauer2016}, regulatory guidance \citep{USFoodandDrugAdministration2019}, and reporting standards \citep{Dimairo2020}.

Cluster randomised trials present a particularly interesting setting for adaptive design. In these experiments, groups such as healthcare facilities, schools, households, or geographical areas, rather than individual participants, are randomly allocated to interventions \citep{Eldridge2012,Turner2021}. They are widely used when an intervention naturally operates at group level or when individual randomisation would lead to contamination between treatment conditions. The resulting design problem differs from that of an individually randomised experiment in several important respects. Outcomes from participants in the same cluster are correlated, so the information provided by additional observations depends on the within-cluster correlation structure. Sample size is also inherently multidimensional: investigators may be able to recruit additional clusters, recruit more participants within existing clusters, extend follow-up, or some combination of these. These components can have very different costs. In longitudinal cluster trials there is an additional design dimension because investigators may determine when clusters switch between treatment conditions. Consequently, both the efficiency and cost of a cluster trial can depend strongly on covariance parameters that are difficult to estimate accurately before the trial begins \citep{Hemming2020,Ouyang2023}.

This uncertainty has motivated a growing literature on adaptive methods for cluster randomised trials. One strand concerns internal pilots and sample-size re-estimation. \citet{Lake2002} proposed re-estimating nuisance parameters during recruitment and updating the required sample size, and \citet{vanSchie2014} developed related methods when further recruitment within existing clusters is possible. More recently, \citet{Sarkodie2024} considered a hybrid Bayesian--frequentist approach in which prior information about the intraclass correlation coefficient (ICC), describing the correlation between individuals in the same cluster, is updated at an interim analysis and used to re-estimate the required sample size. These approaches address an important problem in cluster trials that misspecification of the ICC or other variance parameters at the design stage can lead to substantial under- or over-recruitment.

A second strand concerns sequential monitoring and adaptation of treatment allocation. \citet{Grayling2017} adapted group-sequential error-spending methods to stepped-wedge cluster randomised trials - longitudinal cluster trial designs with an incremental or `staggered' roll out of the intervention over all clusters over the course of the trial - allowing early stopping for efficacy or futility. \citet{Brown2020} subsequently examined information growth and sequential monitoring when the ICC is unknown, while \citet{Hemming2021} discussed the practical and statistical issues involved in interim monitoring of cluster trials. Bayesian approaches have also been developed: \citet{Shen2022} proposed Bayesian group-sequential designs for cluster randomised trials, \citet{Wang2024} considered Bayesian adaptive stopping in stepped-wedge trials, and \citet{Harari2024} described a broader Bayesian framework for adaptive public-health and cluster randomised trials. Adaptation need not be restricted to stopping or sample size. \citet{Grayling2022} considered response-adaptive modification of intervention roll-out in stepped-wedge trials, and \citet{Liu2026} proposed Bayesian response-adaptive randomisation of clusters together with efficacy and futility stopping. Recent work has also examined the operating characteristics of adaptive multi-arm cluster trials when only small numbers of clusters are available \citep{Nolan2025}.

These developments establish that adaptive methods can be used successfully in cluster randomised trials, but they also illustrate a feature of the problem that has received less attention. Existing methods generally focus on a particular type of adaptation, for example, re-estimating one component of sample size, adding sequential stopping to a specified design, or altering treatment allocation. In many cluster trials, however, the second-stage design is itself a multidimensional decision. After an interim analysis, investigators might choose simultaneously how many new clusters to recruit, how much additional recruitment to undertake within existing clusters, how long to continue observation, and how the intervention should be rolled out. These choices can trade statistical information against several different resource constraints. Recruiting a new cluster may, for example, provide more information than recruiting additional participants to an existing cluster but at substantially greater cost. Similarly, in staggered implementation designs, changing the timing of treatment roll-out can alter efficiency without necessarily changing the numbers of clusters or participants \citep{Watson2023}. There is therefore no single scalar ``sample size'' to re-estimate, and designs that minimise expected resource use need not minimise the maximum resources that might ultimately be required.

In this article we develop a two-stage adaptive framework intended for this broader design problem. Our starting point is the combination-test principle used in flexible designs for individually randomised trials \citep{Bauer1994,Proschan1995,Mller2001,Jennison2015}. The principal complication in cluster trials is that the same clusters may contribute observations in both stages, so the evidence accumulated before and after the interim analysis is not marginally independent. We address this by expressing the overall score as the sum of a stage-one marginal score and a stage-two score conditional on the stage-one observations. This provides stage-wise test statistics that can be combined using weights fixed at the planning stage, while allowing the second-stage design to depend flexibly on the interim data.

The framework permits both treatment-effect information and updated estimates of auxiliary covariance parameters to inform the second-stage design. Rather than restricting adaptation to a single sample-size parameter, we optimise over a set of candidate designs that can vary numbers of clusters and participants, trial duration, and intervention roll-out. We distinguish between criteria that minimise expected cost and those that constrain maximum cost, and use Pareto optimality to describe trade-offs when several resource measures are important. We illustrate the approach for a two-stage parallel cluster trial and we apply the framework retrospectively to the E-MOTIVE cluster randomised trial.

\section{A combination score test for cluster trials}
\label{sec:test}
Our analysis is derived from a standard generalised linear mixed model (GLMM) and can be applied to any data that can be described by this model, which is common for cluster trials \citep{Li2021}. The analysis here is based on a two-sided null hypothesis test that the parameter associated with the treatment indicator is zero, as is standard for cluster trials \citep{Hemming2020}.

We subscript matrices and vectors with 1 or 2 for the first and second stage data (or $2|1$ for the second stage conditional on the first), respectively, or use no subscript for the full trial data. The full trial outcome data and the covariance model used to design the trial are represented as:
\begin{equation*}
    \mathbf y = \begin{bmatrix}
        \mathbf y_1 \\ \mathbf y_2
    \end{bmatrix}, \qquad  \Sigma = \begin{bmatrix}
        \Sigma_{11} & \Sigma_{12} \\
        \Sigma_{21} & \Sigma_{22} 
    \end{bmatrix} = W + \Phi D\Phi^\top
\end{equation*}
The far right hand side for $\Sigma$ is the marginal quasi-likelihood expression for the covariance, where $W$ is a diagonal matrix with elements $ \text{diag}\left( \left(\frac{\partial h^{-1}(X\beta)}{\partial (X\beta)}\right)^2 \text{Var}(\mathbf y| \mathbf{u})\right)^{-1}$ \citep{Breslow1993}, which are recognisable as the generalised linear model iterated weights. Here, $h$ is the link function, $\boldsymbol{\eta} = X\boldsymbol\beta + \Phi\mathbf u$ is the linear predictor, and $\text{Var}(\mathbf y| \mathbf{u})$ is the marginal variance conditional on the random effect terms $\mathbf{u} \sim N(0,D)$ and $D$ is the covariance matrix of the random effect terms. To complete our model we also have the design matrix of the fixed and random effects:
\begin{equation*}
    X = \begin{bmatrix}
        \mathbf{x}_1 & X_1 \\ 
        \mathbf{x}_2 & X_2
    \end{bmatrix}, \qquad \Phi = \begin{bmatrix}
        \Phi_1 \\ 
        \Phi_2
    \end{bmatrix}
\end{equation*}
where $\mathbf{x}_1$ and $\mathbf{x}_2$ are the vectors of treatment effect indicators for stages 1 and 2, respectively, with associated regression coefficient $\beta$, and $X_1$ and $X_2$ are the matrices of the other ``nuisance'' fixed effects including the intercept and time period indicators. We then project the treatment indicator on the other variables for stage $t$ to `adjust' for the other fixed effects:
\begin{equation}
    \tilde{\mathbf{x}}_t = \mathbf{x}_t - X_t(X_t^\top \Sigma_{tt}^{-1}X^\top)^{-1}X_t^\top \Sigma_{tt}^{-1}\mathbf{x}_t
\end{equation}

We first define the score for the combined trial data:
\begin{equation*}
    U = \tilde{\mathbf{x}}^\top\Sigma^{-1}(\mathbf y - \boldsymbol \mu)
\end{equation*}
To derive the combination score test we first show that $U = U_1 + U_{2|1}$ where $U_1 = \tilde{\mathbf{x}}_1^\top\Sigma_{11}^{-1}(\mathbf y_1 - \boldsymbol\mu_1)$ and $U_{2|1}$ is defined below. Let:
\begin{equation*}
    \Sigma^{-1} = \begin{bmatrix}
        \Omega_{11} & \Omega_{12} \\
        \Omega_{21} & \Omega_{22}
    \end{bmatrix}
\end{equation*}
with 
\begin{align*}
    \Omega_{11} &= \Sigma^{-1}_{11} - \Sigma^{-1}_{11}\Sigma_{12}S^{-1}\Sigma_{21}\Sigma^{-1}_{11} \\
    \Omega_{12} = \Omega^\top_{21} &= -\Sigma^{-1}_{11}\Sigma_{12}S^{-1} \\
    \Omega_{22} &= S^{-1} \\
    S &= \Sigma_{22} - \Sigma_{21}\Sigma_{11}^{-1}\Sigma_{12}
\end{align*} 
and let $\mathbf r_1 = \mathbf y_1 - \boldsymbol \mu_1$ and $\mathbf r_2 = \mathbf y_2 - \boldsymbol\mu_2$, then 
\begin{equation*}
    \Sigma^{-1} \left(\begin{matrix}
        \mathbf r_1 \\
        \mathbf r_2
    \end{matrix} \right) = 
    \left( \begin{matrix}
        \Sigma^{-1}_{11}\mathbf r_1 -\Sigma_{11}^{-1}\Sigma_{12}S^{-1}(\mathbf r_2 - \Sigma_{21}\Sigma_{11}^{-1}\mathbf r_1) \\
        S^{-1}(\mathbf r_2 - \Sigma_{21}\Sigma_{11}^{-1}\mathbf r_1 )
    \end{matrix} \right)
\end{equation*}
Using that $\mathbf r_2 - \Sigma_{21}\Sigma_{11}^{-1}\mathbf r_1  = \mathbf y_2 - \boldsymbol\mu_{2|1} = \mathbf y_2 - (\boldsymbol \mu_2 - \Sigma_{21}\Sigma_{11}^{-1}(\mathbf y_1 - \boldsymbol \mu_1)) = \mathbf r_{2|1}$ and expanding the formula for $U$:
\begin{align}
    \begin{split}
        U &= \tilde{\mathbf{x}}_1^\top\Sigma^{-1}_{11}\mathbf r_1+(\tilde{\mathbf{x}}_2^\top - \Sigma_{21}\Sigma_{11}^{-1}\tilde{\mathbf{x}}_1)S^{-1}(\mathbf y_2 - \boldsymbol \mu_{2|1}) \\
        &= U_1 + {\tilde{\mathbf{x}}}_{2|1}S^{-1}\mathbf r_{2|1} \\
        &= U_1 + U_{2|1}
    \end{split}
\end{align}
which is the sum of the stage 1 marginal score test statistic $U_1$ and the stage 2 conditional score test statistic ($U_{2|1}$).

We also note that the information about the treatment effect can also be decomposed into a marginal and conditional component. If $\mathcal{I} = \tilde{\mathbf{x}}^\top\Sigma^{-1}\tilde{\mathbf{x}}$ then we define:
\begin{align}
    \begin{split}
        \mathcal{I} &= \mathcal{I}_1 + \mathcal{I}_{2|1} \\
        &= \tilde{\mathbf{x}}_1^\top \Sigma^{-1}_{11} \tilde{\mathbf{x}}_1  + \tilde{\mathbf{x}}_{2|1}S^{-1}\tilde{\mathbf{x}}_{2|1}
        \end{split}
\end{align}

Then the two statistics from stage 1 and 2 are:
\begin{equation}
    Z_1=\frac{U_1}{\sqrt{\mathcal{I}_1}}\text{,  } Z_{2|1} = \frac{U_{2|1}}{\sqrt{\mathcal{I}_{2|1}}}
\end{equation}
Then using the above decompositions the overall combined test statistic is
\begin{equation}
    Z = w_1 Z_1 + w_{2}Z_{2|1}
\end{equation}
where $w_1 = \sqrt{\mathcal{I}_1 / \mathcal{I}}$ and $w_2 = \sqrt{\mathcal{I}_{2|1} / \mathcal{I}}$ with $w_1^2 + w_2^2 = 1$. 

Under the null $Z_1 \sim N(0,1)$ and $Z_{2|1} \sim N(0,1)$ so that $Z$ must also have an $N(0,1)$ distribution. As \citet{Jennison2003} note, the type I error is preserved given the information available at the time of adaptation, but also note that the conditional type I error at each stage needs to be preserved for this to hold. Importantly, this means we must use the weights determined at the initial trial design, rather than modify them after viewing the data. In Section \ref{sec:interim} we distinguish between stage 2 estimated information and stage 1 planning information explicitly.

This establishes the validity of the combination statistic under arbitrary adaptation of the stage 2 design, but it does not by itself determine the critical values applied at each stage. Where the design permits rejection of $H_0$ at the interim analysis, the trial has two opportunities to reject and the stage 1 and stage 2 boundaries must be calibrated jointly, exactly as in a group sequential design \citep{Jennison2000}. The role of the combination test is to ensure that this calibration, performed at the planning stage, remains valid however the stage 2 design is subsequently chosen. We set out the calibration in Section \ref{sec:decision}.

\section{Selecting sample sizes and decision rules}
\label{sec:decision}
We focus on contexts in which the trial design can be modified and recruitment can continue within existing clusters. For example, switching control clusters to the intervention to achieve a more staggered implementation, modifying the trial duration, and so forth. In these cases the structure of the matrices $X_2$, $\Sigma_{12}$, and $\Sigma_{22}$ varies over the second stage design. The question then becomes one of identifying efficient or optimal designs that acheive a desired power for an effect size $\delta$.

\subsection{Small sample bias}
A well recognised issue for the analysis of data from a cluster randomised trial is small sample bias. Parallel trials with fewer than approximately 40 clusters exhibit an inflated type I error for the treatment effect test using a standard z-test. A relatively effective way of controlling the type I error is to use a t-test with degrees of freedom equal to the number of cluster-periods minus the number of fixed effect parameters -- the so-called between-within correction \citep{Hemming2025}. The t-statistics are $T_1$, $T_{2|1}$, and $T$, which are t-distributed statistics with degrees of freedom $\nu_1$, $\nu_{2|1}$, and $\nu$ for the first stage, second stage, and full trial, with non-centrality parameters $\delta\sqrt{\mathcal{I}_1}$, $\delta\sqrt{\mathcal{I}_{2|1}}$, and $\delta\sqrt{\mathcal{I}}$, respectively. To convert to the z-distribution scale for our combination test statistic, where relevant, we specify 
\begin{equation*}
    Z_1 = \Phi^{-1}(F_{t_{\nu_1}}(T_1))
\end{equation*}
and similarly for the other trial stages. 

\subsection{Identifying decision rules}
The decision to be made at interim analysis is whether to continue to stage 2 or to stop for efficacy or futility. If the choice is made to continue then the stage 2 design should be selected to achieve the desired power. We define a grid of possible values of stage 2 parameters $\mathcal{G} = \{(K_2,m_2,...)\}$, which includes $K_2$ the number of new clusters to recruit at stage 2 and $m_2$ the cluster size in stage 2 along with any other parameters determining the design and given observed $z_1$ at the interim analysis. A stage 2 design is then $g \in \mathcal G$.

\subsubsection*{Stopping boundaries}
We permit rejection of $H_0$ at the interim analysis. Let $c_1$ denote the stage 1 efficacy boundary, so that the trial stops and rejects $H_0$ if $|Z_1| > c_1$, and let $c_2$ denote the boundary applied to the combination statistic at the end of stage 2, so that $H_0$ is rejected if $|Z| > c_2$. Under $H_0$ the pair $(c_1, c_2)$ must satisfy
\begin{equation}
\label{eq:alphacal}
    \underbrace{Pr(|Z_1| > c_1)}_{\text{reject at stage 1}} +
    \underbrace{\int_{\mathcal{R}}
      Pr\left(|w_1 z_1 + w_2 Z_{2|1}| > c_2\right)
      \phi(z_1)\,dz_1}_{\text{continue and reject at stage 2}} = \alpha
\end{equation}
where $\mathcal{R} \subseteq [-c_1, c_1]$ is the region over which the trial
proceeds to stage 2, and the inner probability is evaluated with
$Z_{2|1} \sim N(0,1)$. Equation (\ref{eq:alphacal}) is the standard group sequential calibration. What the combination test contributes is that $Z_{2|1} \perp Z_1$ and $w_1^2 + w_2^2 = 1$ hold for \emph{any} stage 2 design $g$, so that boundaries solving (\ref{eq:alphacal}) at the planning stage remain valid under adaptation \citep{Bauer1994}.
 
We take
\begin{equation*}
    c_1 = \frac{z_{\alpha/2}}{w_1}
\end{equation*}
the value at which the stage 1 contribution $w_1 z_1$ to the combination statistic alone attains the fixed-sample critical value, and solve (\ref{eq:alphacal}) numerically for $c_2$. Retaining $c_2 = z_{\alpha/2}$ would not satisfy (\ref{eq:alphacal}) since the conditional probability of rejection given $Z_1 = c_1$ is $\Phi(-2z_{\alpha/2}/w_2) + 1/2 \approx 1/2$, so stopping and rejecting at $|z_1| = c_1$ replaces this with certainty, and the error rate exceeds $\alpha$ by $\int_{|z_1| > c_1}\{1 - CP_0(z_1)\}\phi(z_1)\,dz_1$ with $CP_0$ defined below. 

Evaluating (\ref{eq:alphacal}) requires the region over which the trial may continue to stage 2, and two conventions are available. Under a \emph{non-binding} futility boundary the integral is taken over the whole of $[-c_1, c_1]$, making no allowance for trials that stop early for futility. The interim efficacy stop then spends $2\{1 - \Phi(c_1)\}$ of the error rate and the solution satisfies $c_2 > z_{\alpha/2}$. Under a \emph{binding} boundary the integral is taken over the continuation region alone, crediting the trial with the rejections it forgoes by stopping. The futility boundary then returns error rate as well as consuming it, and where the continuation region is a small part of the null sampling distribution the solution can satisfy $c_2 < z_{\alpha/2}$.

The difference is more consequential here than in a typical individually randomised design, because the futility rule of Section~\ref{sec:decision} can be aggressive. For the design of Section~\ref{sec:example1} the continuation region is approximately $(0, c_1)$, so the non-binding convention declines to credit roughly half of the null sampling distribution. The attained error rate is correspondingly below nominal (i.e. conservative), and since the decision rules are calibrated to a power target the shortfall is absorbed by selecting larger stage 2 designs rather than by a loss of power.

Binding calibration presumes a fixed continuation region, so it requires that the stopping rule be determined at the design stage and applied unchanged at the interim analysis. The boundaries $c_1$ and $c_f$ are fixed when the trial is designed, and the interim estimate $\hat{\theta}_1$ governs only the choice of stage 2 design $g$, not whether the trial continues. Under $H_0$ the conditional probability of rejection at stage 2 does not depend on $g$, since $Z_{2|1} \sim N(0,1)$ whichever design is selected. 

The non-binding convention retains the advantage that it bounds the attained error rate above whatever the futility rule does, including rules that respond to the interim data, and it is the appropriate choice where a data monitoring committee may elect to continue past the futility boundary. Under the binding convention the calibration is circular, since $c_2$ depends on the continuation region, which depends on the decision rules, which are calibrated against a power target computed using $c_2$. We resolve this by iterating from the non-binding solution, which converges in a small number of iterations.
 
\subsubsection*{Conditional power}
Without loss of generality we take $\delta > 0$ as the target effect size for trial design so that $H_1: \beta = \delta$; for $\delta < 0$ replace $Z_1$ by $-Z_1$ and $Z$ by $-Z$ throughout. Two conditional probabilities appear in what follows and it is important to distinguish them. The calibration (\ref{eq:alphacal}) involves the conditional probability of rejecting $H_0$ when $H_0$ holds,
\begin{equation*}
    CP_0(z_1;g) = Pr(|Z| > c_2 \mid Z_1 = z_1, \boldsymbol\beta = 0, g)
\end{equation*}
for stage 2 design $g$, which counts rejection in either direction, since under $H_0$ either is a type I error. Conditional power is the probability of rejecting in the direction for which the trial is powered,
\begin{equation*}
    CP(z_1;g) = Pr(Z > c_2 \mid Z_1 = z_1, \beta = \delta, g)
    = \Phi\left(\frac{w_1 z_1 - c_2}{w_2}
      + \delta \sqrt{\mathcal{I}_{2|1}(g)}\right)
\end{equation*}
Rejection in the opposite direction under $\beta = \delta$ is a directional error rather than power and is excluded; see the subsection on directional decisions below.

The total power for the study is
\begin{equation}
    \text{Power} = \underbrace{Pr(Z_1 > c_1)}_{\text{reject at stage 1}}
    + \underbrace{\int_{c_f}^{c_1} CP(z_1; g)\,\phi(z_1 - \mu_1)\,
      dz_1}_{\text{continue and reject at stage 2}}
\end{equation}
where $c_f$ is the futility boundary. The range of integration differs from that in (\ref{eq:alphacal}) deliberately as the calibration treats futility as non-binding and integrates over the whole of $[-c_1, c_1]$, which is conservative, whereas power must be evaluated over the continuation region alone, since trials stopping for futility cannot reject.

\subsubsection*{Directional decisions}
The definition of $CP$ above counts only rejection in the direction for which the trial is powered. Under $\beta = \delta > 0$ a rejection concluding that the control is superior is a directional error, and although its probability tends to zero as $\mathcal{I}$ increases it is not zero in finite samples so it should not contribute to power. The same consideration applies to the stage 1 term, which counts $Pr(Z_1 > c_1)$ rather than $Pr(|Z_1| > c_1)$.

The stage 2 design is likewise selected by maximising $CP$ rather than the two-sided rejection probability. This is permissible because, as shown in Section \ref{sec:interim}, the type I error rate is preserved for any stage 2 design chosen as a function of the stage 1 data and the selection rule need not be two-sided even though the test is. Restricting selection to the target direction has two consequences. First, $CP(z_1;g)$ is strictly increasing in $z_1$, so the futility rule defines a single threshold $c_f$ and the continuation region is the single interval $(c_f, c_1)$. Summing over both directions instead gives a conditional power that approaches one as $z_1 \to -\infty$, producing an interior futility region and a second continuation region in the lower tail. Such ``inner-wedge'' continuation regions considerably complicate the evaluation of operating characteristics \citep{Wassmer2016}. Second, the trial does not expend stage 2 resource pursuing a rejection in the direction opposite to the one for which it was designed. Evidence of harm is still captured, by the two-sided interim boundary $|z_1| > c_1$ and by the two-sided final test.

The procedure may equivalently be described as two simultaneous one-sided combination tests at level $\alpha/2$, with boundaries $(c_1, c_2)$ and $(-c_1, -c_2)$. Since $c_2 > 0$ these rejection regions are disjoint, so at most one direction can be claimed and the final decision is consistent by construction.

\subsubsection*{Cost penalties}
The determination of the stage 2 decision rules requires maximising the conditional power with a penalty for larger samples \citep{Jennison2015}.  The cost is evidently multi-dimensional, involving clusters, participants, and possibly time periods. For example, we define the stage 2 (proportionate) cost for a parallel trial as
\begin{equation*}
    C(g) = (K_1 + K_2)m_2 + \rho K_2
\end{equation*}
and the stage 1 cost as $C_1 = K_1 m_1 + \rho K_1$, where $K_1 > 0$ is the number of stage 1 clusters per arm, $K_2 \geq 0$ is the number of newly recruited stage 2 clusters per arm, $m_1 > 0$ and $m_2 > 0$ are the average cluster-period sizes, i.e. the number of participants per cluster per time-period, in stages 1 and 2 respectively, and $\rho$ is the ratio of the cost of recruiting a new cluster to that of a new participant. The term $(K_1 + K_2)m_2$ reflects that all clusters, including those carried forward from stage 1, are observed again in stage 2 at cost $m_2$ per cluster, while $\rho K_2$ captures the additional recruitment cost of new clusters.

The decision rules for selecting the stage 2 charactertistics $g$ are identified at the trial design stage. At this stage the actual costs borne by the trial are uncertain as they depend on the interim analyses, and so we aim to minimise some summary of the distribution of possible stage 2 costs. We specify two approaches for selecting the optimal stage 2 design each targeting a different objective. The first, following \citet{Jennison2015}, penalises cost directly:
\begin{equation}
\label{eq:costpen}
    g^*(z_1) = \arg\max_{g \in \mathcal{G}} \left[\mathrm{CP}(z_1;g) - \lambda \cdot C(g)\right]
\end{equation}
where $\lambda > 0$ controls the trade-off between conditional power and cost. In our analyses, we use a bisection search to find the value of $\lambda$ that yields the desired overall power $\pi^*$. Since the Lagrangian optimality condition is independent of $z_1$, a single $\lambda$ applies uniformly, and this criterion implicitly minimises $E[C(g)]$ subject to the power constraint.

The second approach constrains the stage 2 budget directly:
\begin{equation}
\label{eq:budgetcons}
    g^*(z_1) = \arg\max_{g \in \mathcal{G}_{\bar{C}}} \mathrm{CP}(z_1;g), \qquad \mathcal{G}_{\bar{C}} = \left\{g \in \mathcal{G} : C(g) \leq \bar{C}\right\}
\end{equation}
where the cost cap $\bar{C}$ is again found by bisection as the minimum budget achieving $\pi^*$. Since all feasible designs satisfy $C(g) \leq \bar{C}$, the maximum total trial cost is bounded by $C_1 + \bar{C}$, regardless of the interim result, and this criterion minimises $\max C$ subject to the power constraint.

The two criteria produce different decision rules. The cost-penalised criterion allocates larger stage 2 designs where conditional power is most sensitive to additional investment, which can result in high costs near the futility boundary where $z_1$ is small but the density $f(z_1 \mid \delta)$ is low. The budget-constrained criterion imposes a uniform cost ceiling, redistributing power from these unfavourable $z_1$ values to more favourable ones that carry higher density under $H_1$. As a result, the two approaches define a trade-off between expected and maximum cost: the cost-penalised criterion achieves lower $E[C]$ while the budget-constrained criterion achieves lower $C_{\max}$. 

To summarise the decision rules for the stage 2 design :
\begin{enumerate}
    \item \textbf{Efficacy stop:} If $|z_1| > z_{\alpha/2}/w_1$, reject $H_0$ and stop the trial.    
    \item \textbf{Continuation:} For each candidate design $g \in \mathcal{G}$, compute the conditional power $\mathrm{CP}(z_1; g)$ 
    \item \textbf{Optimal design selection:} Choose the stage 2 design according to one of two criteria:
    \begin{enumerate}
        \item \emph{Cost-penalised (minimise expected cost):} Select the design that maximises conditional power minus a cost penalty in Equation (\ref{eq:costpen})
        \item \emph{Budget-constrained (minimise maximum cost):} Select the design that maximises conditional power subject to a cost cap in Equation (\ref{eq:budgetcons})
    \end{enumerate}    
    \item \textbf{Futility stop:} If no design offers positive net benefit, i.e.
    \begin{equation*}
        \max_{g \in \mathcal{G}} \left[\mathrm{CP}(z_1;g) - \lambda \cdot C(g)\right] < 0,
    \end{equation*}
    stop the trial for futility and fail to reject $H_0$.
\end{enumerate}

\subsection{Multi-criteria design choice}
When choosing a two-stage adaptive design in the cluster trial case, we can select both stage 1 and 2 sample sizes at participant and cluster level (as well as the trial design, as we illustrate later). The choice over stage 1 parameters then dictates what the stage 2 design looks like in order to achieve the desired level of power at effect size $\delta$. There are several objectives we may consider for this design decision including: minimise expected total number of participants, minimise worst-case total number of participants, minimise expected total number of clusters, minimise worst-case total number of clusters, minimise expected (proportionate) cost, or maximimise probability of early stopping. These objectives often conflict. Designs with small expected sample sizes can have large maximum sample sizes as aggresive early stopping (or lower powered stage 1 designs) result in a larger stage 2 design when continuing. 

We consider a Pareto optimality approach to selecting potential designs within the design space. For each objective of interest we define an objective function $f_1(g),..,f_p(g)$ for a candidate design defined by parameters $g$. These functions describe the criterion to be minimised such as total clusters or expected costs. Then a design $g$ dominates a design $g'$ if and only if:
\begin{equation*}
    f_k(g) \leq f_k(g')  \quad \forall k \in \{1,...,p \}, \qquad \textrm{and} \qquad  f_k(g) < f_k(g') \quad \text{for at least one }k
\end{equation*}
A Pareto optimal design is a design not dominated by any other, and the Pareto frontier consists of all Pareto optimal designs.

\section{Interim re-estimation of auxiliary parameters}
\label{sec:interim}
The combination test framework permits arbitrary adaptation of the stage 2 design based on stage 1 data without inflating the type I error rate \citep{Bauer1994}. This property extends naturally to re-estimation of auxiliary parameters such as the intracluster correlation coefficient and residual variance at the interim analysis. 

Let $\theta$ denote the vector of variance and correlation parameters governing the covariance structure $\Sigma = \Sigma(\theta)$, and let $\hat{\theta}_1$ denote estimates obtained from the stage 1 data $\mathbf y_1$. At the planning stage, the combination test weights are fixed at
\begin{equation*}
    w_1 = \sqrt{\frac{\mathcal{I}_1(\theta_{\mathrm{plan}})}
    {\mathcal{I}(\theta_{\mathrm{plan}})}}, \qquad
    w_2 = \sqrt{1 - w_1^2}
\end{equation*}
based on a planning value $\theta_{\mathrm{plan}}$. These weights, and hence the efficacy boundary $|z_1| > z_\alpha / w_1$, remain fixed throughout the trial.

At the interim analysis, the stage 2 design is selected by re-evaluating the conditional information $\mathcal{I}_{2|1}(g; \hat{\theta}_1)$ for each candidate design in the grid $g \in \mathcal{G}$ using the updated estimates. The conditional power is then calculated instead using $\hat{\theta}_1$. The optimal stage 2 design is then selected by applying the cost-penalised or budget-constrained criterion to (\ref{eq:cp_updated}).

Interim estimates of the correlation parameters are subject to considerable imprecision. In a two-stage design the stage 1 data typically comprise a small number of clusters, and the sampling distribution of the estimated ICC is wide, skewed, and has non-negligible mass at the boundary. This imprecision is difficult to accommodate analytically within the optimisation of the stage 2 design, since the selected design is a step function of $\hat \theta_1$ over a discrete grid, and small changes in the estimate can move the selection discontinuously. The type I error rate is unaffected. Imprecision therefore costs efficiency rather than validity. 

For this reason we regard a simulation study as an essential component of adaptive cluster trial design, and one that should report operating characteristics that directly capture variability. The distribution of the realised sample size and cost rather than their expectations alone, the distribution of the selected stage 2 designs, and the discrepancy between anticipated and realised conditional power, alongside the type I error rate and power. We provide an example below.

\section{Example: Two-stage adaptive parallel trial}
\label{sec:example1}

\subsection{Adaptive trial design}
We consider an example of an adaptive parallel cluster randomised trial. The primary outcome is continuous and the target standardised effect size is $\delta = 0.25$ with desired power of 80\%. The ICC is 0.05 and cluster autocorrelation coefficient (CAC) of 0.8. We set $\rho = 30$, i.e. it costs 30 times more to recruit a cluster than an individual. The cost minimising non-adaptive design in this context includes 18 clusters per arm and 20 participants per cluster-period for a power of 81.6\%. We assume two trial periods to mirror the correlation structure of the adaptive design. This design has 1,440 total participants and a proportional (i.e. up to the constant of proportionality $\lambda$) cost of 2,640.

\begin{figure}
    \centering
    \includegraphics[width=\linewidth]{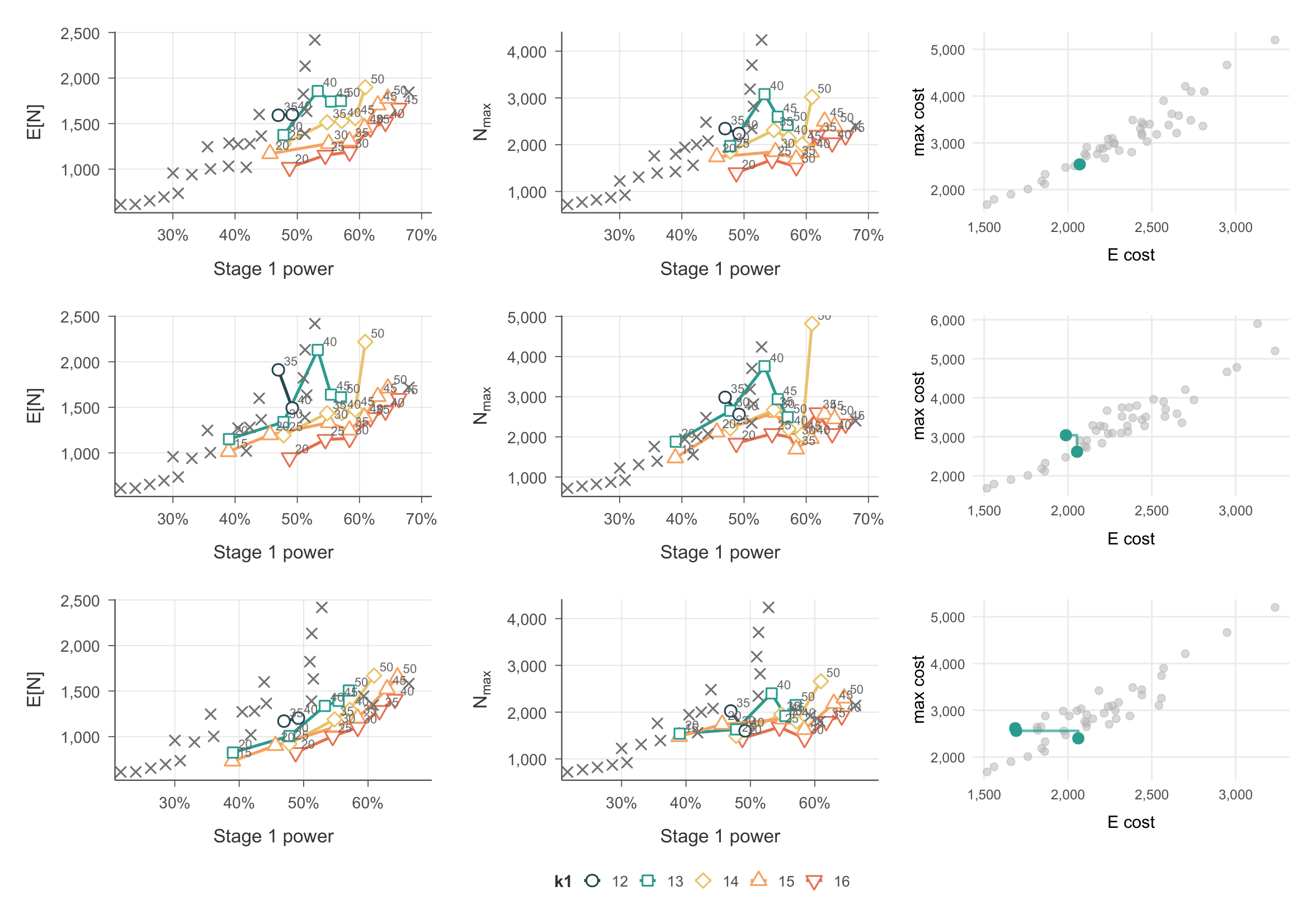}
    \caption{Comparison of stage 1 alternative designs for the adaptive parallel trial example, grey crosses indicate designs with insufficient total power. Left column -- expected total number of participants versus stage 1 power; middle column -- maximum number of participants versus stage 1 power; right column -- maximum versus expected costs with Pareto optimal design highlighted; top row -- cost penalised designs with non-binding futility stopping; middle row -- budget constrained designs with no futility stopping; bottom row -- cost penalised designs with binding futility stopping.}
    \label{fig:parallel1}
\end{figure}

\begin{figure}
    \centering
    \includegraphics[width=\linewidth]{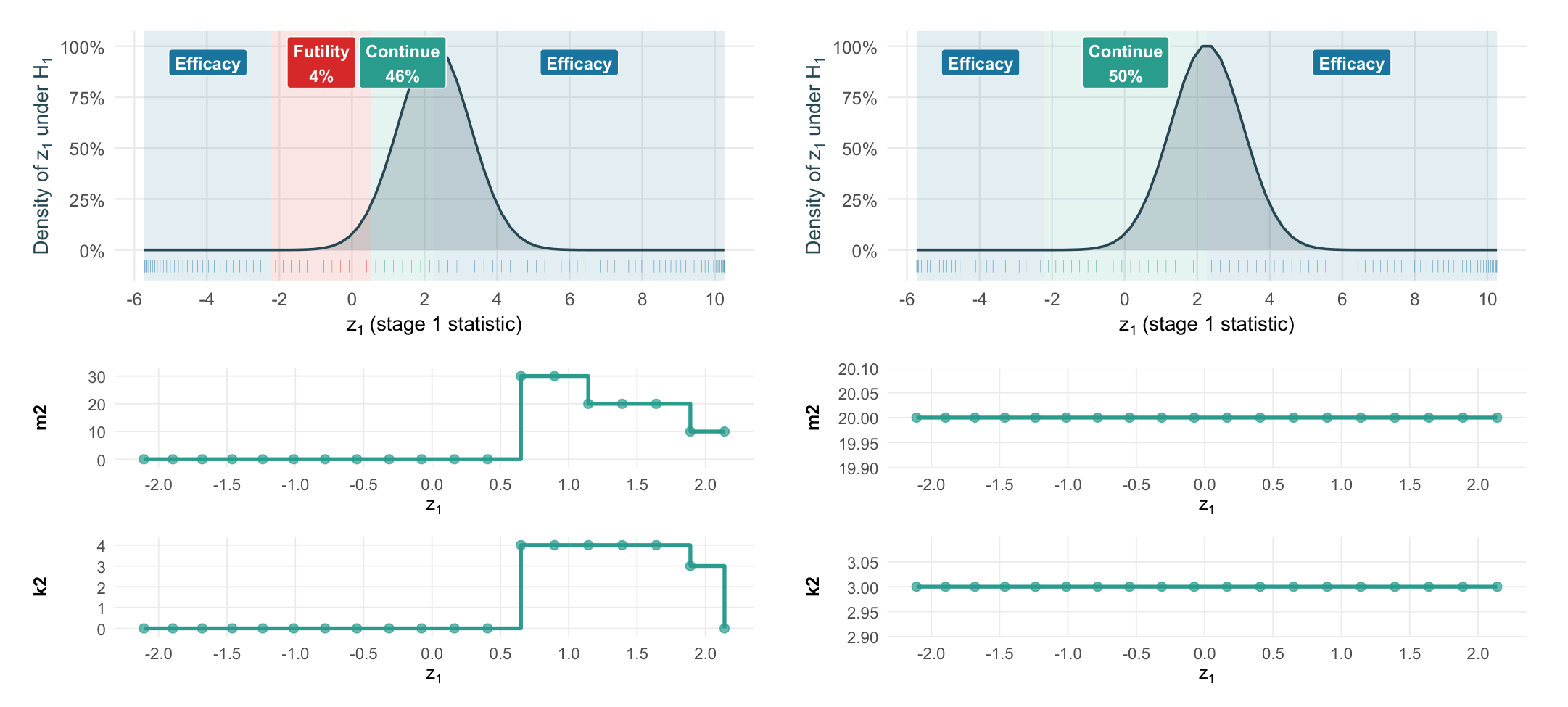}
    \caption{Decision rules for stage 2 number of new clusters (k2) and stage 2 cluster period size (m2) for all stage 2 clusters for the adaptive parallel trial example for a stage 1 test statistic $z_1$, with distribution of stage one test statistic under $H_1$}
    \label{fig:parallel2}
\end{figure}

We identify a Pareto optimal adaptive design where we aim to minimise both the expected and maximum total proportional cost. We consider both optimisation strategies, the cost penalised and the budget constrained approaches. We do not add updates to the auxiliary parameters, which is considered in the next section. We first examine the design space including stage 1 designs with between 12 and 16 clusters per arm, and up to 50 participants per cluster. A maximum of 4 additional clusters per arm and 100 participants per cluster was set. Figure \ref{fig:parallel1} shows the expected and maximum total numbers of participants and costs. We require the stage 1 design to have approximately $>40$\% power to achieve an overall power of 80\%. The budget constrained optimisation achieves a similar expected sample size and cost and marginally smaller maximum cost. Binding futility stopping further reduces the expected sample size. In both cases, the expected trial cost is around 20\% lower, with the maximum costs being 47\% higher than the non-adaptive design. Figure \ref{fig:parallel2} shows the decision rules and probabilities for both approaches.

We selected a stage 1 sample size of 15 clusters per arm of size 20 with an expected and maximum number of participants of 1,198 and 2,120, respectively. Adding a binding constraint reduces the expected and maximum sample sizes to 902 and 2,120, respectively. For the budget constrained optimisation the expected and maximum number of participants are 1,169 and 2,120, respectively. There was no planned futility stopping with the budget constrained method.

\subsection{Simulation-based comparison}
\label{sec:simulation}
The expected sample size and operating characteristics of the design provided above are under an assumed known set of design parameters, including the ICC. As we described in Section \ref{sec:interim}, simulation-based comparisons may be important to quantify the effect of uncertainty in estimates of auxilliary parameters. Here, we present a simulation-based comparison of the design rules for the parallel cluster trial. The trial was designed at an assumed ICC of 0.05, we simulate the 1,000 trials with ICCs of 0.01, 0.05, and 0.10, to represent both under- and over-estimation at the design stage, each under both $H_0$ and $H_1$. We applied each set of rules at the interim stage in each simulated trial example. For the cost penalised approach we also compared binding and non-binding rules; under the budget constrained approach there is no stopping for futility so the rules coincide.

Table \ref{tab:sim_results} shows the results of the simulation-based analysis. The type I error is controlled under all the values of ICC and different types of rules under $H_0$. The role of binding for futility stopping affected the power and expected sample size with the expected cost minimisation rules. The expected sample size with non-binding futility stopping was 1,213, around 16\% smaller than the non-adaptive design, while with binding rules it was 1,046. Both values closely match those from the earlier analysis. All designs had the same maximum sample size of 2,120. The additional risk of type I error through continuation led to more conservative rules in stage 2. However, where the ICC was smaller than planned expected sample sizes dropped to 600 - 750 versus the main trial. Sample sizes equivalently inflated for an ICC of 0.10 although given the bounds on stage 2 expansion, the target power was not achieved and was between 60-75\%.

\begin{sidewaystable}[]
    \centering
    \begin{tabular}{ccc|cccc|c}
    \toprule
    Hypothesis & True ICC & Futility rule & Futility & Efficacy & Continue & Reject rate & $E[N]$ (sd(N)) \\
    \midrule
    \multicolumn{8}{c}{Cost penalised} \\
    \midrule
     $H_0$ & 0.01 & Binding & 0.51 & 0.02 & 0.47 & 0.049 & 746 (173)   \\
     $H_0$ & 0.05 & Binding &  0.52 & 0.02 & 0.46 & 0.050 & 910 (490) \\
     $H_0$ & 0.10 & Binding & 0.52 & 0.02 & 0.46 & 0.057 & 1,213 (719)  \\
     $H_1$ & 0.01 & Binding & 0.00 & 0.63 & 0.37 & 0.920 & 718 (169)  \\
     $H_1$ & 0.05 & Binding & 0.02 & 0.42 & 0.57 & 0.808 & 1,046 (567)  \\
     $H_1$ & 0.10 & Binding & 0.04 & 0.28 & 0.68 & 0.682 & 1,544 (711)  \\
     \midrule
     $H_0$ & 0.01 & Non-binding & 0.92 & 0.02 & 0.05 & 0.025 & 633 (182)   \\
     $H_0$ & 0.05 & Non-binding & 0.46 & 0.02 & 0.52 & 0.030 & 1,218 (700)  \\
     $H_0$ & 0.10 & Non-binding & 0.06 & 0.02 & 0.92 & 0.049 & 1,675 (615)  \\
     $H_1$ & 0.01 & Non-binding & 0.15 & 0.64 & 0.22 & 0.825 & 688 (199)  \\
     $H_1$ & 0.05 & Non-binding & 0.06 & 0.42 & 0.52 & 0.773 & 1,180 (672)  \\
     $H_1$ & 0.10 & Non-binding & 0.02 & 0.25 & 0.73 & 0.571 & 1,674 (681)  \\
     \midrule
    \multicolumn{8}{c}{Budget constrained} \\
    \midrule
     $H_0$ & 0.01 & Both & 0.00 & 0.01 & 0.99 & 0.048 & 959 (200)   \\
     $H_0$ & 0.05 & Both & 0.00 & 0.02 & 0.98 & 0.052 & 1,441 (535)  \\
     $H_0$ & 0.10 & Both & 0.00 & 0.01 & 0.99 & 0.041 & 1,712 (564)  \\
     $H_1$ & 0.01 & Both & 0.00 & 0.64 & 0.36 & 0.895 & 735 (220)  \\
     $H_1$ & 0.05 & Both & 0.00 & 0.44 & 0.56 & 0.753 & 1,224 (657)  \\
     $H_1$ & 0.10 & Both & 0.00 & 0.27 & 0.73 & 0.602 & 1,660 (681)  \\
     \bottomrule
    \end{tabular}
    \caption{Results from 1,000 trial simulations of the adaptive parallel trial under different true ICCs and decision rules}
    \label{tab:sim_results}
\end{sidewaystable}

\section{Applied Example: E-Motive}
\label{sec:emotive}
The E-MOTIVE trial was a cluster randomised trial comparing a multi-component clinical intervention for postpartum hemorrhage in patients having a vaginal delivery \citep{Gallos2023}. The primary outcome was a binary indicator for several maternal outcomes. The original trial was designed as a parallel trial with baseline. The parameters used in the original sample size calculation were an ICC of 0.02, CAC of 0.97, a baseline prevalence of 2\% and a target effect size of -0.6 percentage points or a 30\% relative risk reduction with power 90\% and type I error of 5\%.

The trial had a sample of 80 clusters and enrolled 210,132 patients. The target sample size in the baseline and post-intervention periods was 1,350. We used the same design parameters to identify a two-stage adaptive approach for this trial. We considered stage 1 designs with and without a baseline period of data collection with variable numbers of clusters per arm, and cluster sizes in baseline and post-intervention phases. The designs could then enroll more clusters and increase cluster sizes in the post-interim analysis phase. We used an expected cost minimisation criteria with a ratio of 200:1 for clusters:participants for the cost function and non-binding futility stopping rules. 

\begin{figure}
    \centering
    \includegraphics[width=\linewidth]{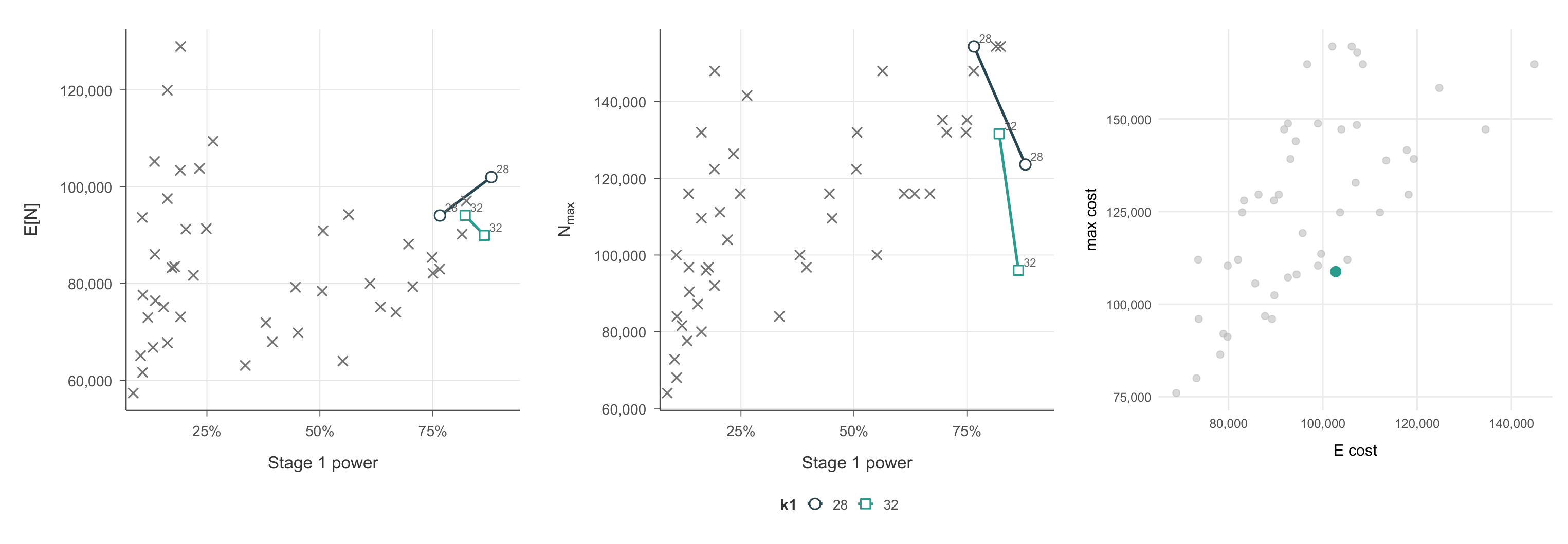}
    \caption{Comparison of the expected and maximum total number of patients, and of expected and maximum costs of the range of stage 1 designs considered for E-MOTIVE with futility non-binding cost minimisation rules.}
    \label{fig:emotive1}
\end{figure}

\begin{figure}
    \centering
    \includegraphics[width=\linewidth]{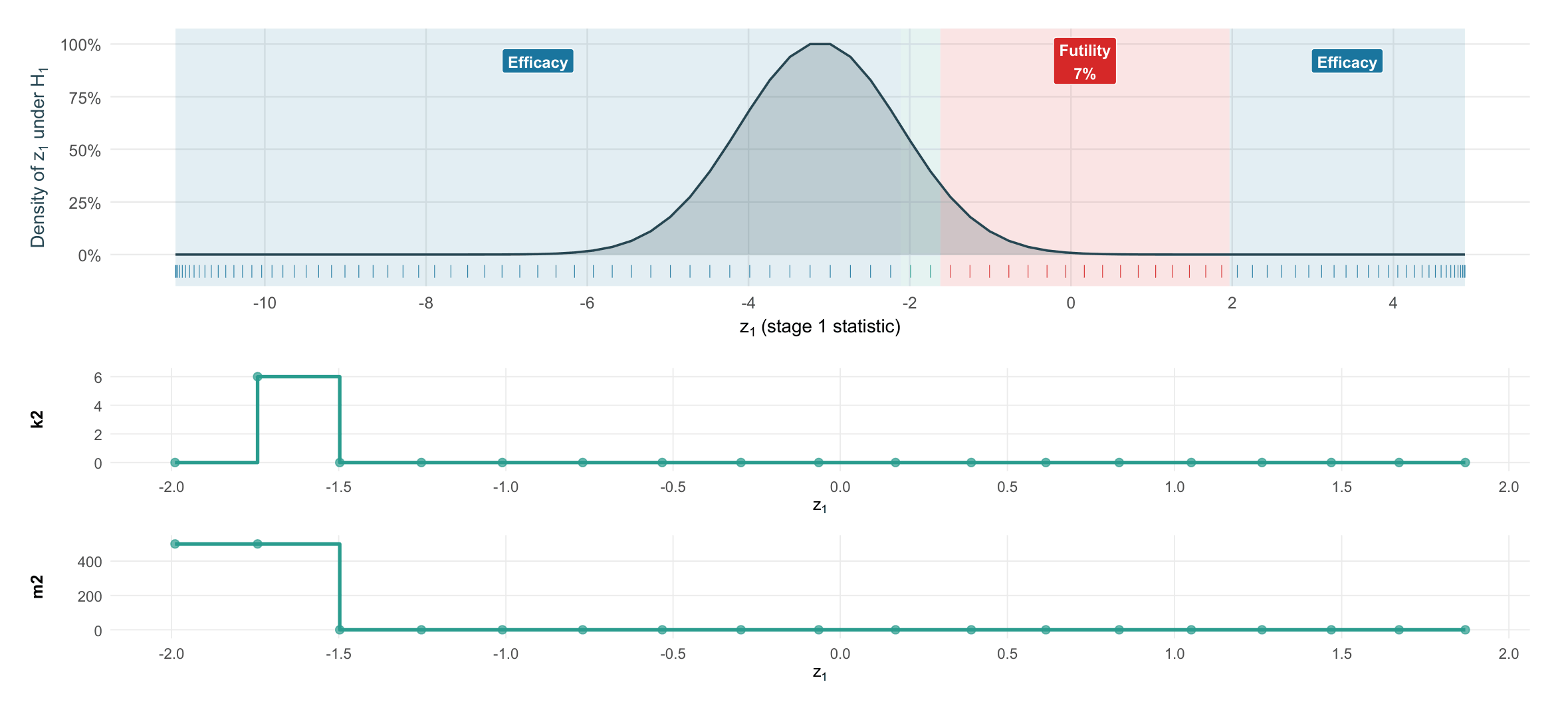}
    \caption{Decision rules for the two-stage adaptive E-MOTIVE trial design under an expected cost minimisation rule. k2 is the number of newly recruited stage 2 clusters and m2 the stage 2 recruitment in clusters.}
    \label{fig:emotive2}
\end{figure}

Figure \ref{fig:emotive1} shows the comparisons of the different designs examined. One design was identified as being Pareto optimal when comparing maximum and expected costs. This design enrolled 32 clusters per arm with up to 500 baseline observations and 900 post-intervention observations in stage 1. The decision rules for this design are shown in Figure \ref{fig:emotive2}. Up to six clusters could be enrolled in stage 2 with expanded patient recruitment of up to 500 patients per cluster in the worst case scenario. The z-statistic cut-off for efficacy was -2.082 in the first stage and -2.059 in the second stage under the non-binding futility stopping rule. For the analysis we followed the same statistical specification as the original trial. We sampled patients and clusters from the original trial dataset to recreate the stage 1 analysis according to their order of entry into the original trial. 

Table \ref{tab:emotive} shows the results of the hypothetical interim analysis. The stage 1 z-statistic was -5.22, which would have led to a decision to stop the trial for efficacy. The total sample size at the point of the interim analysis was 64 clusters and 79,964 patients, with some clusters not achieving the target enrollment. At interim we included 20\% fewer clusters and more than 60\% fewer patients that the original trial. However, the early stopping would have precluded the trial from investigating longer term effects and sustainability of the intervention. In this case, would could redistribute the sample size or budget to later time periods to consider the question of temporally heterogeneous effects. 

\begin{table}[]
    \centering
    \begin{tabular}{c|c}
    \toprule
    Stage 1 statistic & Value \\
    \midrule
       Odds ratio (95\% CI) & 0.52 (0.40, 0.66) \\
       z-statistic & -5.22 \\
        ICC (95\% CI) & 0.013 (0.009, 0.017) \\
        CAC & 0.56\\
        \bottomrule
    \end{tabular}
    \caption{Stage 1 interim results for the E-MOTIVE trial re-analysis}
    \label{tab:emotive}
\end{table}

\section{Conclusions}
\label{sec:discussion}
Adaptive methods have a long history in clinical trial design. By integrating interim analyses into the course of a trial, decisions can be made to improve the efficiency of the design, remove ineffective treatments, or stop for efficacy. All of these options may reduce the costs of running the study and expose fewer patients to an experiment. Adaptive designs are therefore potentially attractive to funders, patients, and trialists. Cluster randomised trials may particularly benefit from adaptive, flexible designs as their power and sample size is often determined by multiple uncertain correlation parameters, and they have potential flexibility in intervention implementation schedules, follow-up durations, and sample sizes at multiple levels. There have been some examples of adaptive cluster trial approaches in the context of comparing multiple treatments and dropping ineffective alternatives (e.g. \citet{Choko2019}). However, there are few examples of adaptive design cluster trials based on unblinded interim analyses.

We have presented a method to extend two-stage flexible design methods to the cluster trial setting. The combination test methods are general within the class of GLMMs and can be accommodated to more complex correlation structures, cluster size heterogeneity, or more complex trial designs. The methods allow adaptation across a range of trial characteristics. Estimation of inferential statistics including p-values and confidence intervals is also relatively straightforward and can be based on the final pooled information matrix using the planned weights. Interim re-estimation of the auxiliary correlation parameters can improve the efficiency of the design, but the estimates available at an interim analysis in a cluster trial are typically imprecise, and the benefit is not guaranteed. Our simulations indicate that re-estimating the ICC can significantly reduce expected sample sizes if it is smaller than planned. We did not see much difference in expected sample sizes between the two cost functions and the magnitude of any advantage depended on whether the futility stopping rules were binding or not at the planned sample size. We therefore recommend that adaptive cluster trial designs be evaluated by simulation across a range of plausible correlation structures before the trial is initiated, with attention to the variability of the operating characteristics, not only their expected values, as well as the role of different rules.

Two-sided testing is the norm in cluster randomised trials, but it interacts with adaptive designs in ways that require care. Because the combination weights are fixed at the planning stage, the direction in which the test rejects need not coincide with the direction indicated by the pooled estimate when the realised stage 2 information departs substantially from the planned value \citep{Wassmer2016}. A dual test, rejecting in a given direction only when the pooled statistic agrees in sign, removes this possibility and can only reduce the type I error rate, so requires no recalibration. In our simulations the condition was never invoked, reflecting the bounded stage 2 designs considered here, but it may be more relevant where much larger sample size increases are permitted.

Extensions to these methods include multi-stage designs with multiple interim analyses. These extensions would require the specification of independent marginal and conditional test statistics for each stage. Although we illustrate that one can achieve sizeable reductions in expected sample size with a two-stage design. An important consideration for complex cluster trials may also be that there are multiple effects of interest, such as an immediate or time-averaged effect, and a temporally heterogeneous or sustained effect. Other examples may include estimating both direct and indirect effects of a vaccine, where different designs may provide better efficiency for different effects. In these cases, one may re-design the trial at the interim stage to focus on the effect with the greatest uncertainty and balance trial resources to answer multiple questions. Future research will focus on these multi-objective designs.

\bigskip
\begin{center}
{\large\bf SUPPLEMENTARY MATERIAL}
\end{center}

\begin{description}

\item[R-package:] R-package acrt is available online doi: 10.5281/zenodo.18875018 with reproduction materials for the analyses in this article and general functionality for the methods described in this article.

\end{description}

\begin{center}
{\large\bf DATA AVAILABILITY STATEMENT}
\end{center}
The example in this study use synthetic data available in the linked R package above. The E-Motive data were used with permission from the original study authors \citet{Gallos2023} and are available on request from that study's authors.

\bibliography{adapt}
\end{document}